\documentclass[twocolumn,showpacs,preprintnumbers,amsmath,amssymb]{revtex4}

\usepackage{graphicx}
\usepackage[latin1]{inputenc}
\usepackage{amsmath}
\usepackage{amsfonts}
\usepackage{amssymb}
\usepackage{amsthm}
\usepackage{amsxtra}
\usepackage{fancyhdr}
\usepackage{mathrsfs}
\newcommand{\be}{\begin{equation}}
\newcommand{\ee}{\end{equation}}
\newcommand{\ben}{\begin{eqnarray}}
\newcommand{\een}{\end{eqnarray}}

\begin{document}

\title{Separability Criteria and Entanglement Measures 
for Pure States of $N$ Identical Fermions}

\author{A. R. Plastino}
\affiliation{Instituto Carlos I de F\'{\i}sica Te\'orica y
Computacional, Universidad de Granada, 18071-Granada, Spain, EU}
\affiliation{National University La Plata, UNLP-CREG, C.C.
727, 1900 La Plata, Argentina} 
\author{D. Manzano}
\affiliation{Instituto Carlos I de F\'{\i}sica Te\'orica y
Computacional, Universidad de Granada, 18071-Granada, Spain, EU}
\affiliation{Departamento de F\'{\i}sica At\'omica, Molecular, y
Nuclear, Universidad de Granada, 18071-Granada, Spain, EU}

\author{J. S. Dehesa}

\affiliation{Instituto Carlos I de F\'{\i}sica Te\'orica y
Computacional, Universidad de Granada, 18071-Granada, Spain, EU}
\affiliation{Departamento de F\'{\i}sica At\'omica, Molecular, y
Nuclear, Universidad de Granada, 18071-Granada, Spain, EU}

\date{\today}

\begin{abstract}
The study of the entanglement properties of systems of $N$ fermions
has attracted considerable interest during the last few years. 
Various separability criteria for pure states of $N$ identical 
fermions have been recently discussed but, excepting the case 
of two-fermions systems, these criteria are difficult to implement 
and of limited value from the practical point of view. Here we 
advance simple necessary and sufficient separability criteria 
for pure states of $N$ identical fermions. We found that to be identified as
separable a state has to comply with one single identity involving
either the purity or the von Neumann entropy of the single-particle 
reduced density matrix. These criteria, based on the verification 
of only one identity, are drastically simpler than the criteria 
discussed in the recent literature. We also derive two inequalities 
verified respectively by the purity and the entropy of the single 
particle, reduced density matrix, that lead to natural entanglement 
measures for $N$-fermion pure states. Our present considerations  
are related to some classical results from the Hartree-Fock 
theory, which are here discussed from a different point of view in
order to clarify some important points concerning the separability 
of fermionic pure states.

\end{abstract}


\maketitle

Entanglement constitutes an essential ingredient of the quantum
mechanical description of the physical world \cite{BZ06,S07}. It
is also a physical resource with important technological
implications \cite{NC00}. A fundamental first step in the study of
the entanglement properties of a given class of quantum systems is
the establishment of appropriate separability criteria. That is,
to establish criteria that enables us to tell if a given quantum
state is separable or entangled. A good separability criterion, besides
its obvious importance as a tool for determining the presence or
absence of entanglement, is also relevant as the possible basis of
quantitative measures of entanglement. An appropriate measure of
the deviation of the actual properties of a given quantum state
from those required by the separability criterion may provide a
valuable estimation of the amount of entanglement exhibited by 
that state.

The study of the entanglement features of systems consisting of
$N$ identical fermions has attracted the attention of many
researchers in recent years
\cite{AFOV08,LV08,NV07,BBB07,GM04,GMW02,ESBL02,LZLL01,BPCP08,OSTS08}.
Entanglement between fermionic particles has been studied in
connection with various physical scenarios. To mention just a few
examples, researchers have recently investigated entanglement in
two-electrons atomic states \cite{AM04}, entanglement between
pairs of electrons in a conducting band \cite{NV07}, entanglement
dynamics in two-electrons scattering processes \cite{BBB07}, 
and the role of entanglement in time-optimal evolutions of 
fermionic systems \cite{BPCP08,OSTS08}, among many others. 
Appropriate separability criteria (and entanglement measures) 
for pure states of two identical fermions have been recently 
derived (using the Schmidt decomposition) and applied to
the study of various physical systems and processes
\cite{NV07,BBB07,GM04}. Alas, the aforementioned derivations of
separability criteria cannot be extended to situations involving
more than two fermions because in such cases the Schmidt
decomposition doesn't exist. 

Some separability criteria for more than two fermions have 
been proposed in the recent quantum information literature, 
but they are difficult to implement in practice and exhibit 
a growing degree of complexity when one increases the number 
of particles of the system or the dimensionality of the 
single-particle Hilbert space. 
The necessary and sufficient criterion introduced by Eckert,
Schliemann, Bruss, and Lewenstein \cite{ESBL02} (from now on ESBL)
is based on a projection operator acting upon an $N$-fermion state
and resulting in an $(N-1)$-fermion state. This operator
depends on an arbitrary single-particle state $|a\rangle$. The
ESBL criterion says that a pure $N$-fermion state $|\Psi \rangle$
has Slater rank one (that is, it is a separable state) if and only
if the result of applying the projector operator on $|\Psi\rangle$
is, {\it for any single-particle state $|a\rangle$}, either equal
to an $(N-1)$-fermion state of Slater rank $1$ or  equal to zero.
The ESBL separability criterion has been recently hailed
\cite{AFOV08} as the main result known so far on necessary and
sufficient separability criteria for $N$-fermion pure states. The
ESBL criterion certainly is of considerable relevance from the
fundamental  and conceptual points of view, but it is of little
practical use. To check if a given state $|\Psi\rangle$ fulfils
the ESBL criterion is, in general, basically as difficult as the
original problem of finding out if $|\Psi \rangle$ has Slater rank
equal to $1$ or not. The ESBL criterion can be iterated $N-2$
times, leading to a chain of separability tests eventually ending
with a separability test to be performed on a two-fermion state.
However, this procedure does not reduce the difficulty of the
criterion, since each link in the aforementioned chain involves a
relation that has to be checked for an arbitrary single particle
state $|a\rangle$ \cite{ESBL02}. A different approach employing
sophisticated techniques from algebraic geometry has been advanced
in \cite{LV08}. According to this proposal, however, to be
identified as separable a quantum $N$-fermion state has to comply
with several relations (that is, not just with one identity as in
the criterion proposed by us), their number increasing with the
number of fermions in the system.

The aim of the present work is to derive two inequalities 
verified, respectively, by the purity $Tr \left( \rho_r^2 \right)$
and the von Neuman entropy $-Tr\left( \rho_r \ln \rho_r \right)$
of the single particle reduced density matrix $\rho_r$ of
an $N$-fermions pure state. These inequalities lead to 
simple separabilty criteria and suggest practical entanglement 
measures. These separability criteria turn out to be closely 
related to some previous results from the theory of Hartree-Fock 
wave functions that, even though themselves constituting useful 
necessary and suficient separability criteria, doesn't seem to 
have been recognized as such in the recent literature. Our derivations 
are different from (and simpler than) the ones followed in 
the aforementioned works on the Hartree-Fock wave functions. 
Moreover, our developments clarify why those previous results
have not been believed to provide sufficient separability
criteria for $N$-fermions states.

Let us consider a system consisting of a {\it constant} number $N$
of identical fermions with a single particle Hilbert space of
dimension $D$, with $N\le D$ (if $N>D$ it is not possible to
construct an antisymmetric $N$-fermion state). A pure state of
such a system is separable (that is, non-entangled) if it has
Slater rank equal to one \cite{AFOV08}. That is to say, the state
is non entangled if it can be expressed as a single Slater
determinant,

\be \label{slater}
a_{i_1}^{\dagger} ... a_{i_N}^{\dagger} |0\rangle,
\ee

\noindent where $a_i^{\dagger}$ are fermionic creation operators
acting upon the vacuum state $|0\rangle $ and leading to an orthonormal
basis $\{|i\rangle = a_i^{\dagger}|0\rangle \}$ of the single-particle
Hilbert space. A pure state of the
$N$-fermion system that cannot be written in the above way has a
finite amount of entanglement. Correlations between the $N$
fermions that are due solely to guarantee the antisymmetric
character of the fermionic states do not contribute to the state's
amount of entanglement \cite{GM04,GMW02,ESBL02}. There are profound 
physical reasons for this. On the one hand, these correlations 
(exhibited by states with Slater rank 1) can't be used as a 
resource to implement non-classical information transmission 
or information processing tasks \cite{ESBL02}. On the other hand,
the non-entangled character of states represented by one Slater
determinant is consistent with the possibility of associating complete 
sets of properties to both parts of the composite system
(see \cite{GM04,GMW02} for an interesting, detailed discussion of this approach).

When discussing the entanglement properties of systems of $N$ 
identical fermions the relevant group of ``local transformations'' 
is isomorphic to the group $SU(D)$ of (special) unitary transformations 
acting on the $D$-dimensional single-particle Hilbert space \cite{ESBL02}. 
Given a transformation $U \!\! \in \!\! SU(D)$ the corresponding ``local transformation''
acts on a general $N$-fermions state according to
$\sum w_{i_1, \ldots, i_N}\,
a_{i_1}^{\dagger} \ldots a_{i_N}^{\dagger} |0\rangle \rightarrow
\sum w_{i_1, \ldots, i_N}\,
{\tilde a}_{i_1}^{\dagger} \ldots {\tilde a}_{i_N}^{\dagger} |0\rangle$, 
where ${\tilde a}_i^{\dagger} |0\rangle =  |{\tilde i}\rangle$ and
$U| i \rangle  = |{\tilde i} \rangle$, ($i=1, \ldots, D$). The set
of non-entangled fermionic states is closed under the action of
these ``local transformations''. Furthermore, the entanglement measures
that we are going to consider in this work are invariant under 
those transformations.

A simple illustration of the fact that the correlations associated 
with a fermionic state of Slater rank $1$ cannot be used as a resource 
for quantum information tasks is provided by a two-electrons system with
a four dimensional relevant single-particle Hilbert space \cite{ESBL02}.
Let us assume that the relevant single-particle Hilbert space admits a 
basis of the for $\{|\phi_1\rangle |+\rangle,|\phi_1\rangle |-\rangle,
|\phi_2\rangle |+\rangle,|\phi_2\rangle |-\rangle\}$,
where $|\phi_{1,2}\rangle$ are two spatial wave functions and $|\pm\rangle$
corresponds to the spin degree of freedom. The two electrons can be treated as
effectively distinguishable entities if they are spatially localized. 
This can occur if the moduli of $\langle {\bf r} | \phi_1 \rangle$ and
$\langle {\bf r} | \phi_2 \rangle$ are non-overlapping. The single 
particle basis can then be partitioned between two agents (Alice and Bob),
$\{|\phi_1\rangle |+\rangle,|\phi_1\rangle|-\rangle\}$ being the basis of
Alice's space and $\{|\phi_2\rangle |+\rangle,|\phi_2\rangle | -\rangle\}$
the basis of Bob's space. Under these circumstances, a state
of the form $\frac{1}{\sqrt{2}}\Bigl(|\phi_1\rangle| +\rangle \otimes 
|\phi_2\rangle | -\rangle - |\phi_2\rangle | -\rangle \otimes 
|\phi_1\rangle| +\rangle \Bigr) $
given by a single Slater determinant (and describing two particles 
localized in different spatial regions) effectively behaves
as the non-entangled (in the usual sense) state $|\phi_1 \rangle|+ \rangle_A \otimes
|\phi_2 \rangle|- \rangle_B$ describing two distinguishable objects ($A$ and $B$).
On the other hand, a state describing two localized electrons that cannot
be cast as one single Slater determinant effectively behaves as 
an entangled state (in the standard sense corresponding to 
distinguishable subsystems) that is useful for performing non-classical 
information related tasks (see \cite{ESBL02} for a more detailed discussion).

The amount of entanglement associated with an $N$-fermion state corresponds,
basically, to the quantum correlations exhibited by the state on
top of the minimum correlations needed to comply with the
antisymmetric constraint on the fermionic wave function.  Note
that here we are considering entanglement between {\it particles},
and not entanglement between {\it modes} (see \cite{BCW07} for a
comprehensive discussion of entanglement between modes).

 Given a single particle orthonormal basis $\{|i\rangle
 =a_i^{\dagger}|0\rangle, \,\, i=1, \ldots, D\}$,  any pure
 state of the $N$-fermion system can be expanded as,

\be \label{fermistate}
\left|\Psi\right>=\sum_{i_1, \ldots, i_N = 1}^D
w_{i_1, \ldots, i_N}\,
a_{i_1}^{\dagger} \ldots a_{i_N}^{\dagger} |0\rangle,
\ee

\noindent
where the complex coefficients $w_{i_1, \ldots, i_N}$
are antisymmetric in all indices and comply with the
normalization condition

\be \label{norma}
\sum_{i_1, \ldots, i_N = 1}^D
|w_{i_1, \ldots, i_N}|^2 \, = \, \frac{1}{N!}.
\ee

\noindent The single-particle reduced density matrix $\rho_r$
associated with the $N$-fermion pure state (\ref{fermistate}) has
matrix elements,

\be \label{redensimat}
\langle i | \rho_r | j \rangle
\, = \, \frac{1}{N} \,
\langle \Psi |a_j^{\dagger} \, a_i | \Psi \rangle,
\ee

\noindent
where the factor $1/N$ guaranties that $\rho_r$
is normalized to unity,

\be \label{rhonormuno}
Tr \rho_r \, = \, 1.
\ee

Let $F_i\equiv \left<i\right|\rho_r\left|i\right>$ denote
the diagonal elements of $\rho_r$. After some algebra
it is possible to verify that,

\be \label{rediagonal}
F_i=  \sum_{\substack{(i_1, \ldots, i_n)\\ i_1<i_2<...<i_n}}
(N!)^2 \, |w_{i_1, \ldots, i_N}|^2
\, f_i^{(i_1, \ldots, i_N)},\,\,\,\,\,
 i=1,\ldots, D.
\ee

\noindent
where

\begin{equation} \label{probaf}
f_i^{(i_1, \ldots, i_N)}=\left\{
\begin{array}{cc}
\frac{1}{N},& \,\, {\rm if }\quad i\in (i_1, \ldots, i_N), \\
0&\quad {\rm otherwise.}\\
\end{array}
\right.
\end{equation}

\noindent
Note that the sum in (\ref{rediagonal}) has only
$\binom{D}{N}=\frac{D!}{N!(D-N)!}$ terms
because it doesn't run over all the $D^N$ possible
$N$-uples $(i_1, \ldots, i_N)$; it runs only
over the $\binom{D}{N}$ $N$-uples whose indices
are all different and listed in increasing order.
Thus, the vector ${\bf F}$
(with components $\{F_i, \,\,\, i=1,\ldots, D \}$)
can be expressed as a linear combination of the $\binom{D}{N}$
vectors  ${\bf f^{(i_1, \ldots, i_N)}}$ (with components
$\{f_i^{(i_1, \ldots, i_N)}, \,\,\, i=1,\ldots, D\}$).
Each one of these vectors
has $D$ components, $N$ of them being equal to $1/N$ and the
rest equal to zero. To simplify notation it is convenient
to introduce a single global label $k$, $1\le k\le \binom{D}{N}$,
to characterize the coefficients
$ (N!)^2 \, |w_{i_1, \ldots, i_N}|^2 $
and the vectors ${\bf f^{(i_1, \ldots, i_N)}}$. Equation
(\ref{rediagonal}) can then be recast in a more compact way as,

\be \label{redig}
F_i=\sum_{k=1}^M d_k f_{ik},
\ee

\noindent
where $M=\binom{D}{N}$
and the identifications

\begin{align}
(N!)^2 \, |w_{i_1, \ldots, i_N}|^2 & \to d_k \nonumber  \\
f_i^{(i_1, \ldots, i_N)}& \to f_{ik}
\end{align}

\noindent
have been made. We have $0\le d_k \le 1,\,\,\, (1\le k \le M)$,
$0\le f_{ik} \le 1,\,\,\, (1\le k \le M; \, 1\le i \le D)$,
and,

\be \label{lasnormas}
\sum_{k=1}^Md_k = 1;      \,\,\,\,\,\,\,\,
\sum_{i=1}^D f_{ik} = 1;  \,\,\,\,\,\,\,\,
\sum_{i=1}^D f_{ik}^2 = \frac{1}{N}.
\ee

\noindent
The vector ${\bf F}$ and each
of the vectors ${\bf f_k}$ can be regarded as
properly normalized probability distributions,
and the vector ${\bf F}$ is a {\it convex}
linear combination of the vectors ${\bf f_k}$.

Let us now consider the sum of the squares of the
components of the vector ${\bf F}$,

\begin{align} \label{demo1}
\sum_{i=1}^D F_i^2
=&\sum_{i=1}^D\left\{ \left( \sum_{k=1}^M d_k^2 f_{ik}^2  \right)+\right. \nonumber\\
&\qquad\left. 2 \left( \sum_{k<k'} d_k d_{k'} f_{ik} f_{ik'} \right) \right\}\nonumber\\
=&\sum_{i=1}^D\left\{ \left( \sum_{k=1}^M d_k
\left(  1-\sum_{k'\ne k}d_{k'}\right) f_{ik}^2\right) +\right.\nonumber\\
&\qquad\left.  2 \left(\sum_{k<k'}d_k d_{k'} f_{ik} f_{ik'}\right) \right\}\nonumber\\
=&\sum_{i=1}^D\left\{ \left( \sum_{k=1}^M d_k f_{ik}^2  \right)-
\left( \sum_{k\ne k'} d_k d_{k'} f_{ik}^2 \right)+\right.\nonumber\\
 &\qquad 2 \left. \left(\sum_{k<k'} d_k d_{k'} f_{ik} f_{ik'} \right)\right\}\nonumber\\
=&\sum_{i=1}^D\left\{\!\! \left( \sum_{k=1}^M d_k f_{ik}^2\right)\right.\nonumber\\
 &\qquad -\left.
\sum_{k<k'}\!\! d_k d_{k'} \! \left( f_{ik}^2 \! +
\! f_{ik'}^2 \! - \! 2 f_{ik}f_{ik'}\right) \! \right\}\nonumber \\
=&\left\{\sum_{k=1}^M d_k \left( \sum_{i=1}^D f_{ik}^2\right)\right\}\nonumber\\
 &\qquad -\left\{\sum_{k<k'} d_k d_{k'}\sum_{i=1}^D \left(f_{ik}-f_{ik'}\right)^2\right\}.
\end{align}

\noindent
Since $\sum_{i=1}^D f_{ik}^2=\frac{1}{N}$ for all $k$, it follows from
(\ref{demo1}) that,

\begin{align} \label{desigual1}
\sum_{i=1}^D  \left<i\left|\rho_r\right|i\right>^2=&\sum_{i=1}^D F_i^2
\nonumber\\
=&\frac{1}{N} - \underbrace{\sum_{k<k'}d_k d_{k'}
\sum_{i=1}^D\left(f_{ik}-f_{ik'} \right)^2}_{\ge 0}\nonumber\\
\le& \frac{1}{N}.
\end{align}

\noindent 
The inequality in (\ref{desigual1}) can also be obtained applying 
Jensen inequality to the square of the right hand side of (\ref{redig})
and taking into account the first and the third equations in
(\ref{lasnormas}).

The only way for the equality sign to hold in (\ref{desigual1})
is to have one of the $d_k$ equal to $1$ and the rest
equal to $0$, meaning that there is only one term in the original
expansion for $\left|\Psi\right>$. This implies that
$\left|\Psi\right>$ has Slater rank one, and can thus be expressed
as one single Slater determinant. Since we didn't impose any
restriction on the single-particle basis $\{|i\rangle\}$, equation
(\ref{desigual1})  holds for any such a basis. In particular, it
holds for the eigenbasis of the single-particle reduced
statistical operator $\rho_r$, implying that

\be \label{desigual2}
Tr\left(\rho^2_r \right) \le \frac{1}{N}.
\ee

\noindent It is easy to see that when the Slater rank of the
$N$-fermions state $|\Psi \rangle$ is one we have
$Tr\left(\rho^2_r \right)= \frac{1}{N}$. On the other hand,
$Tr\left(\rho^2_r \right)= \frac{1}{N}$ implies that there exists
a single-particle basis for which the equal sign holds in
(\ref{desigual1}), implying in turn that the state under
consideration has Slater rank $1$ and it is then separable.

Summing up, the following double implication obtains,

\be \label{eureka1}
\left|\Psi\right> {\rm has\;Slater\;rank\;one}
\iff
Tr\left( \rho_r^2 \right)=\frac{1}{N}.
\ee

\noindent
In other words, {\it a pure state of $N$ identical fermions
is separable if and only if the purity of the reduced
single-particle density matrix is equal to $1/N$}.

It is possible to formulate a separability criterion
equivalent to (\ref{eureka1}) in terms of the von
Neumann entropy of the single particle density matrix
$\rho_r$.  Let us consider the Shannon entropies
of ${\bf F}$ and ${\bf f_k}$ (regarded as probability
distributions),

\be \label{nielsen}
S[{\bf F}] = -\sum_{i=1}^D F_i \ln F_i; \,\,\,\,
S[{\bf f_k}] = -\sum_{i=1}^D f_{ik} \ln f_{ik}
\ee

\noindent
Using the concavity property of the Shannon entropy
\cite{CT91}, it follows from (\ref{redig}) that,

\be \label{convex}
S[{\bf F}] \ge \sum_{k=1}^M d_k S[{\bf f_k}] = \ln N,
\ee

\noindent where the inequality reduces to an equality if and only
if all the probability vectors ${\bf f_k}$ appearing in the sum in
the middle term in (\ref{convex}) are equal to each other. This
can only happen if one of the $d_k$'s is equal to $1$ and the rest
are equal to zero. That is, it can happen only if the $N$-fermion 
state can be written as a single Slater determinant. Equation 
(\ref{convex}) holds for any single-particle basis $\{|i\rangle \}$. 
In particular, it holds for the eigenbasis of $\rho_r$, which leads 
to

\be \label{ineqineq}
S\left[\rho_r \right] \ge \ln N.
\ee

\noindent It is plain that an $N$-fermion pure state with Slater
rank one leads to a single-particle reduced density matrix
verifying $S\left[ \rho_r \right] = \ln N$. Conversely, the
relation $S\left[ \rho_r \right] = \ln N$ implies that there
exists a single-particle basis such that $-\sum \langle i | \rho_r
|i\rangle \ln\langle i | \rho_r |i\rangle=  S[{\bf F}] = \ln N$
which, as we have already seen, implies that the $N$-fermion pure
state can be written as a single Slater determinant and,
consequently, describes a separable state. Summarizing,

\be \label{eureka2}
\left|\Psi\right> {\rm has\;Slater\;rank\;one}
\! \iff \!
-Tr\!\!\left( \rho_r \ln \rho_r \right)
\! = \! \ln N.
\ee

A particular instance of the separability
criterion (\ref{eureka2}), corresponding
to  systems of two identical fermions,
has already been discussed by Ghirardi and
Marinatto in \cite{GM04}. The derivation of
the $N\!=\!2$ case of (\ref{eureka2}) given by
Ghirardi and Marinatto is based upon the
Schmidt decomposition for systems of
two fermions. Unfortunately, the Schmidt
decomposition does not exist when $N\ge3$
 and, consequently, the developments presented
in \cite{GM04} cannot be extended to situations
involving systems of three or more identical
fermions. Our present treatment, besides providing
a necessary and sufficient separability
criterion valid for arbitrary values of the
number $N$ of particles, is also of interest
as yielding an alternative way of obtaining
the $N=2$ criterion without recourse to the
Schmidt decomposition.

The necessary and sufficient condition
for separability $Tr[\rho_r^2] =1/N$ 
is closely related to the condition

\be \label{sueco}
\rho_r^2 \, = \, \frac{1}{N} \rho_r
\ee

\noindent
that the single particle reduced density matrix 
has to verify if the global wave function
can be expressed as a Slater determinant.
Condition (\ref{sueco}) has been discussed 
in the past in the context of atomic physics  
\cite{RH94,L55} and actually constitutes a classicall 
result from the theory of the Hartree-Fock 
approximation. However, the relevance of 
condition (\ref{sueco}) as a useful separability
criterion for $N$-fermions pure states has not 
been properly appreciated within the field of 
quantum entanglement theory. In fact, condition 
(\ref{sueco}) has been in the recent literature
regarded as not providing a necessary separability 
criterion. In fact, in connection with $N$-fermions
states leading to a reduced density matrix
verifying (\ref{sueco}) it has been recently
stated that ``{\it ... a wave function of this 
kind can in general not be written as a single Slater 
determinant constructed from orthogonal states}'' 
\cite{ESBL02}.  As we are going to show next, 
our present results show in a direct and manifest 
way that the alluded wave functions can indeed 
be written as a single Slater determinant 
constructed from orthogonal states (that is, they
have Slater rank $1$).

Note that condition (\ref{sueco}) is not, by itself,
equivalent to either the relation (\ref{eureka1})
or to the entropic relation (\ref{eureka2}).  It 
is plain that a density matrix $\rho_r$ complying
with (\ref{sueco}) must necessarily verify relations
(\ref{eureka1}) and (\ref{eureka2}). However, the reciprocal
implication doesn't hold. A density matrix verifying
(\ref{eureka1}) (or verifying (\ref{eureka2})) does not 
necessarily fulfil (\ref{sueco}). For instance, 
if $\rho_r$ has eigenvalues 
$(\frac{1}{2},\frac{1}{2\sqrt{2}}, \frac{1}{2\sqrt{2}}, 0)$
we have that $Tr[\rho_r^2]=\frac{1}{2}$ but 
$\rho_r^2 \ne \frac{1}{2}\rho_r$. However, it 
follows from our proof of the separability conditions
(\ref{eureka1}) and (\ref{eureka2}) that either 
of the relations $Tr[\rho_r^2]=\frac{1}{N}$ or 
$S[\rho_r] = \ln N$, {\it together with the additional
information that the single particle statistical
operator $\rho_r$ comes from an $N$-fermion pure state},
guarantee that equation (\ref{sueco}) is verified
(since in that case we have an equality in equation
(\ref{desigual1}) and the global state must have Slater 
rank $1$, implying that the only possible values for 
the eigenvalues of $\rho_r$ are $1/N$ and $0$).
In other words, {\it in the special case of statistical
operators $\rho_r$ that are reduced single particle
matrices arising from an $N$-fermion state} we
have the double implication

\be
Tr\left( \rho_r^2 \right)=\frac{1}{N}
\iff
\rho_r^2 \, = \, \frac{1}{N} \rho_r.
\ee

\noindent
Consequently, and contrary to some current beliefs,
equation (\ref{sueco}) does provide a necessary
and sufficient criterion for separability of
$N$-fermion states.

Finally, note that on the light
of the separability criteria (\ref{eureka1}) and (\ref{eureka2})
it is reasonable to regard the differences

\ben \label{medent}
{\cal E}_L \, = \, \frac{1}{N} \, - \,
Tr\left(\rho_r^2 \right) \cr
{\cal E}_{VN} \, = \, S\left[\rho_r \right]
\, - \, \ln N,
\een

\noindent as measures of the amount of entanglement exhibited by a
pure state of a system of $N$ identical fermions. The quantities
(\ref{medent}) have already been proposed as measures of
entanglement for fermions (particularly for two-fermion systems. 
See the excellent review \cite{AFOV08} on entanglement 
in many-particle systems) but our present results lend 
considerable further support to that proposal, because we 
now know with certainty that the measures (\ref{medent}) 
are non-negative quantities that vanish if and only if 
the fermionic pure state under consideration is separable.
In the particular case of systems of two fermions with
$D=4$ the quantity $4{\cal E}_L$ reduces to the entanglement
measure (usually referred to as squared concurrence)
studied in \cite{ESBL02} (see also \cite{BPCP08}).

Summing up, we have derived a couple of inequalities
involving respectively the purity and the von Neumann 
entropy of the single particle, reduced density matrix
$\rho_r$ of an $N$-fermion pure state. These inequalities 
lead directly to simple and practical (necessary and sufficient) 
separability criteria based on the verification of one single 
identity. These criteria are drastically simpler than others 
that have been considered (for $N>2$) in the recent 
literature. Moreover, the aforementioned inequalities also 
suggest two practical measures of entanglement for fermionic 
pure states. In the particular case of $N=2$ the separability
criteria discussed by us reduce to the criteria derived
in \cite{GM04} (see also \cite{NV07}) by recourse to the 
fermionic Schmidt decomposition.

\begin{acknowledgments}
The authors gratefully acknowledge the Spanish MICINN grant FIS2008-02380 and
the grants FQM-2445, FQM-481, and FQM-1735 of the Junta de Andaluc\'{\i}a. 
They belong to the Andalusian research group FQM-207. D.M acknwoledges 
the corresponding FPI scholarship of the Spanish Ministerio de 
Ciencia e Innovaci\'on.
\end{acknowledgments}


\end{document}